\documentstyle[12pt]{article}

\def\C{{\bf C}}
\def\tl{{Temperley-Lieb }}
\def\tlm{{Temperley-Lieb-Martin }}
\def\tr{{\rm tr}}
\def\proof{\goodbreak\noindent{\bf Proof.\quad}}
\def\endproof{{\ $\Box$}\bigskip }
\newtheorem{prop}{Proposition}[section]
\newtheorem{lemma}[prop]{Lemma}
\newtheorem{thm}[prop]{Theorem}
\newtheorem{df}[prop]{Definition}
\newtheorem{rk}[prop]{Remark}
\newtheorem{cor}[prop]{Corollary}

\title{Representation-theoretic derivation of the
Temperley-Lieb-Martin algebras}
\author{
Tomasz Brzezi\'nski\thanks{Address after 1st October 1995:
University of Cambridge, DAMTP, Cambridge CB3 9EW, U.K.} and
Jacob Katriel\thanks{
Permanent address: Department of Chemistry,
Technion - Israel Institute of
Technology, Haifa, Israel.} \\
{\small \sl Physique Nucl\'eaire Th\'eorique et Physique
Math\'ematique} \\
{\small \sl Universit\'e Libre de Bruxelles} \\
{\small \sl Campus de la Plaine CP229,}
{\small \sl B1050 Bruxelles, Belgium}}
\date{ }
\begin{document}
\setcounter{page}{1}

\maketitle

\vspace*{5 cm}

\begin{abstract}
Explicit expressions for the Temperley-Lieb-Martin algebras,
{\it i.e.}, the
quotients of the Hecke algebra that admit only representations
corresponding to Young diagrams with a given maximum number of
columns (or rows), are obtained, making explicit use of the Hecke
algebra representation theory. Similar techniques are used to
construct the algebras whose representations do not contain
rectangular subdiagrams of a given size.
\end{abstract}

\vspace{3cm}

{\bf PACS:} 0210, 0220, 0550.

\newpage
\baselineskip 21pt
\section{Introduction}

The Temperley-Lieb algebra plays an important role in a wide range
of areas of mathematics and physics. It grew out of a study of
relations between the percolation and colouring problem \cite{tl},
and has since been used in  studies of integrable models in
statistical mechanics \cite{Martinb},
von Neumann algebras \cite{Jones83},
representations of braid groups, and knot and link
invariants \cite{Jones87,Good,Kauffman}.

The Temperley-Lieb algebra $TL\sb n(q)$ is known to be a quotient of
the Hecke algebra $H\sb n(q)$ to an algebra, whose irreducible
representations are classified by Young diagrams with at most two
columns or, equivalently, two rows \cite{Jones87,Good}.
$H_n(q)$ is defined as a free unital associative algebra generated by
$g\sb 1, \ldots, g\sb{n-1}$ subject to the relations
\begin{equation}
\label{equation:Hecke}
\begin{array}{ll}
g_i^2=(q-1)g_i+q &\;\;\; i=1,\, 2,\, \cdots,\, n-1 \\
g_ig_{i+1}g_i=g_{i+1}g_ig_{i+1} &\;\;\; i=1,\, 2,\, \cdots,\, n-2 \\
g_ig_j=g_jg_i &\;\;\; {\mbox{if }} |i-j|\geq 2
\end{array}
\end{equation}
where $q$ is a complex number. It may be considered as
a deformation of the group algebra $\C S\sb n$ of the symmetric group
$S\sb n$. In the limit $q=1$
the $g\sb i$ are identified with transpositions $(i,i+1)$. For generic
values of $q$ the representation theory of $H\sb n(q)$  closely
resembles the representation theory of $S\sb n$. In particular,
irreducible representations of $H\sb n(q)$ can be labelled by
Young diagrams with $n$ boxes \cite{Wenzl,Vershik}.

To obtain the defining relations of $TL\sb n(q)$
we first define
\begin{equation}
e_j=\frac{g_j+1}{q^{\frac{1}{2}}},
\label{equation:defe}
\end{equation}
rewrite the first and third of relations (\ref{equation:Hecke})
in terms of the generators $e\sb j$
\begin{eqnarray}
&&e_j^2=(q^{1/2}+q^{-1/2})e_j\; ,\\
&&e_ie_j=e_je_i \qquad {\mbox{if }} |i-j|\geq 2 ,
\end{eqnarray}
and, instead of the second of relation (\ref{equation:Hecke}) impose
the condition
\begin{equation}
\label{equation:bra}
e_je_{j\pm 1}e_j-e_j =0.
\end{equation}
Thus, the \tl algebra $TL\sb n(q)$ is the free associative algebra
generated by $e\sb 1, \ldots e\sb{n-1}$, subject to the relations
(3), (4), and (\ref{equation:bra}).

The \tl algebra has been applied to the analysis of integrable
models with the quantum group $U\sb q(su(2))$ symmetry \cite{Pasquier}.
In order to analyse integrable models
such as spin chains or diffusion-reaction processes \cite{Alcaraz},
which have $U\sb q(su(N))$ symmetry, one has to consider
multi-column (or multi-row)
generalisations of the \tl algebra, that have recently been
introduced by Martin \cite{Martinb}.
These \tlm algebras are defined as the
quotients of the Hecke algebra $H\sb n(q)$ that admit only irreducible
representations described by Young diagrams with at most $N$
columns. Such an algebra for $N=3$ was also considered by
Sochen \cite{Sochen}. Another generalisation of the \tl algebra, that
excludes Young diagrams
containing a rectangular Young subdiagram with specified
numbers of both rows and columns, was considered by
Martin and Rittenberg \cite{Martin}. In this paper we give a
simple derivation of these algebras, making explicit use of
the representation theory of the Hecke algebra $H_n(q)$, and,
in particular, of the properties of the Murphy operators
\cite{Dipper,Murphy}.
We find that with a certain choice of generators the defining
relations of the multi-column \tl algebra are
obtained as higher order iterates of the \tl braiding
relation, eq. (\ref{equation:bra}).

Our paper is organised as follows. In Section~2 we describe some
elements of the representation theory of Hecke algebras, focusing on
the construction of projection operators. In Section~3 we state and
prove the main result concerning the structure of the \tlm algebras.
In Section~4 we discuss the structure of a
double quotient of the Hecke algebra which leads to
an algebra whose Young diagrams do not contain a rectangle
of a given shape. In Section~5 we make some concluding remarks.

We assume that $q$ is a real positive number.

\section{Projection operators in the Hecke algebra $H_n(q)$}

In this section we define and discuss the basic properties of certain
projection operators in the Hecke algebra $H\sb n(q)$. We start,
however, recalling some elementary facts about the Young diagrams.
We denote by
$\Gamma_n\equiv [\lambda_1,\, \lambda_2,\, \cdots,\, \lambda_k]$
the Young diagram of $n$ boxes arranged into $k$ rows of
respective lengths $\lambda\sb 1 \geq \lambda\sb 2 \geq\ldots \geq
\lambda\sb k$.

\begin{df}
The Young diagram
$\Gamma\sb n\equiv [\lambda_1, \, \lambda_2, \, \cdots , \, \lambda_k]$
contains the Young diagram
$\Gamma^{\prime}\sb m\equiv [\lambda_1^{\prime}, \, \lambda_2^{\prime},
\, \cdots , \, \lambda_{\ell}^{\prime}]$
if $k\geq\ell$ and
$\lambda_i\geq\lambda_i^{\prime},\, \; \; i=1,\, 2,\, \cdots,\,\ell$.
\label{young.contain}
\end{df}

The following lemma and corollary are immediate consequences of the
Littlewood-Richardson rule for the outer product
\cite{Hamermesh}.

\begin{lemma}
 Any Young diagram obtained as the outer product
of two Young diagrams $\Gamma_n$ and $\Gamma_m$ contains both.
\label{lemma.young.product}
\end{lemma}

\begin{cor}
A Young diagram with no more than $\ell$ columns
(rows) could only be obtained as a direct product of Young
diagrams with no more than $\ell$ columns (rows).
\label{cor.product}
\end{cor}

Both Lemma~\ref{lemma.young.product} and Corollary~\ref{cor.product}
will be used in the proof of the main result
in Section~3.

\begin{df}
The $q$-content of the $j$'th box in the
$i$'th row of a Young diagram is $q[j-i]_q$, where
$[k]_q\equiv \frac{q^k-1}{q-1}=1+q+\cdots+q^{k-1}$.
\label{def.qcontent}
\end{df}

In the analysis of the structure of Hecke algebras an important role
is played by the Murphy operators \cite{Dipper,Murphy} $L\sb p$, given
by
\begin{eqnarray*}
L_p &=& g_{p-1}+
\frac{1}{q} g_{p-2}g_{p-1}g_{p-2}
+ \frac{1}{q^2} g_{p-3}g_{p-2}g_{p-1}g_{p-2}g_{p-3}
\\ & & {\phantom{space here}} +\cdots
+ \frac{1}{q^{p-2}} g_1g_2\cdots g_{p-2}g_{p-1}g_{p-2}\cdots g_2g_1
\; ;\;\; \; \; \; \; \;
p=2,\, 3,\, \cdots,\, n. \nonumber
\end{eqnarray*}
Any two Murphy operators commute with one another.
The symmetric polynomials in the Murphy operators span the
center of the Hecke algebra.
A state labelled by the sequence of Young diagrams
$\Gamma_2\subset\Gamma_3\subset\cdots\subset\Gamma_n$
is an eigenstate of all the Murphy operators
$L_2,\, L_3,\, \cdots,\, L_n$.
The eigenvalue of $L_i$  corresponding to this state
is the $q$-content of the box that has
been added to $\Gamma_{i-1}$ to obtain $\Gamma_i$.

The fundamental invariant of $H_n(q)$, $\sum_{i=2}^n L_i$,
has been shown to fully characterise its irreducible representations
\cite{Katriel1},
and could therefore be used to construct projection
operators onto subspaces consisting of irreducible representations
with any desired specification.
However, using the properties of the Murphy operators
such projection operators can be written down in an even
simpler form.
This is particularly simple for the one-dimensional single
row $[n]$ or single column $[1^n]$ irreducible representations.
Take $[n]$ for example and
note that
it is the only irreducible representation of $H_n(q)$ for which
the box at position
$(2,1)$ does not exist.  The $q$-content of this box is $-1$.
Thus, for all irreducible representations but $[n]$ one of the
Murphy operators must assume the eigenvalue $-1$. Therefore,
${\tilde{C}}_n\equiv\prod_{i=2}^n (L_i+1)$ vanishes on
all irreducible representations except $[n]$.
The eigenvalues of the Murphy operators corresponding to the various
boxes of $[n]$
are $q,\, q+q^2,\, q+q^2+q^3,\, \cdots,\, q+q^2+\cdots+q^{n-1}$,
respectively, so that the operator ${\tilde{C}}_n$  assumes the
value $[n]_q!$ where  $[i+1]_q!=[i]_q![i+1]_q$.
Thus, the normalised projection operator on the single row
irreducible representation of $H_n(q)$ is
$$C_n=\prod_{j=2}^n \frac{L_j+1}{[j]_q}\; .$$
This projection operator can be written as a sum over the $n!$
reduced words that furnish a basis of $H_n(q)$. Explicitly the first
two operators come out as
\begin{eqnarray*}
C_2 &=& \frac{1}{[2]_q}(1+g_1)\, ,\\
C_3 &=& \frac{1}{[3]_q!}(1+g_1+g_2+g_1g_2+g_2g_1+g_1g_2g_1)\, .
\end{eqnarray*}

Since $C_n$ is manifestly symmetric in the Murphy operators,
it is central in $H_n(q)$. Being a projection operator, it is
idempotent. One important consequence is that for an arbitrary
polynomial $F(g_1,g_2,\cdots,g_{n-1})$ in the generators of $H_n(q)$
the identity
\begin{equation}
\label{equation:identity}
F(g_1,g_2,\cdots,g_{n-1})C_n=F(q,q,\cdots,q)C_n
\end{equation}
is satisfied in any irreducible representation of $H\sb n(q)$.
This follows from the fact that on the single-row
irreducible representation all $g_i$ are represented by $q$, and
on all other irreducible representations
both sides of eq. (\ref{equation:identity}) vanish.

For our construction of quotients of the Hecke algebra $H\sb n(q)$
it is  necessary to consider the Hecke subalgebras
generated by sets of consecutive generators
$g_i,\, g_{i+1},\,\cdots ,\, g_{i+\ell-2}$, which are isomorphic
with $H_{\ell}(q)$. We shall denote such algebras by
$H_{\ell}^{(i)}(q)$, where the superscript specifies the
generator with the lowest index. Within those subalgebras
the Murphy operators are
$$L_j^{(i)}=g_{j+i-2}+\frac{1}{q}g_{j+i-3}g_{j+i-2}g_{j+i-3}+\cdots
+\frac{1}{q^{j-2}}g_ig_{i+1}\cdots g_{j+i-2}\cdots g_{i+1}g_i
\; ; \; j=2,\, 3,\,\cdots,\, \ell$$
and the projection operator onto the one-row irreducible representation
of $H\sb{\ell}\sp{(i)}(q)$ is
$$C_{\ell}^{(i)}=\prod_{j=2}^{\ell}\frac{L_j^{(i)}+1}{[j]_q}\; .$$
We often suppress the superscript when the lowest generator is $g_1$.

We now derive a recurrence relation for the single-row
projection operators.
First we use the idempotency of $C_{i+1}$ and the fact that
it commutes with $L_{i+2}$ to note that
\begin{equation}
\label{equation:ab}
C_{i+2}=C_{i+1}\frac{L_{i+2}+1}{[i+2]_q}=
C_{i+1}\frac{L_{i+2}+1}{[i+2]_q}C_{i+1}.
\end{equation}
Next, using eq. (\ref{equation:identity}) we obtain
\begin{equation}
\label{equation:i1}
C_{i+1}\frac{L_j^{(2)}+1}{[j]_q}C_{i+1}=C_{i+1}
\; ; \;\;\; j=2,\,3,\,\cdots ,\, i,
\end{equation}
\begin{equation}
\label{equation:i2}
C_{i+1}L_{i+2}C_{i+1}=C_{i+1}g_{i+1}C_{i+1}[i+1]_q,
\end{equation}
and
\begin{equation}
\label{equation:i3}
C_{i+1}L_{i+1}^{(2)}C_{i+1}=C_{i+1}g_{i+1}C_{i+1}[i]_q\; .
\end{equation}
 From eqs. (\ref{equation:i2}) and (\ref{equation:i3}) it follows that
\begin{equation}
\label{equation:i4}
C_{i+1}\frac{L_{i+2}+1}{[i+2]_q}C_{i+1}=
\frac{[i+1]_q^2}{[i]_q[i+2]_q}C_{i+1}\frac{L_{i+1}^{(2)}+1}{[i+1]_q}
C_{i+1}-\frac{q^i}{[i]_q[i+2]_q}C_{i+1}.
\end{equation}
Substitution of eq. (\ref{equation:i4}) in eq. (\ref{equation:ab})
and use of eq. (\ref{equation:i1}) yields
$$C_{i+2}=\frac{[i+1]_q^2}{[i]_q[i+2]_q}
C_{i+1}^{(1)}C_{i+1}^{(2)}C_{i+1}^{(1)}
-\frac{q^i}{[i]_q[i+2]_q}C_{i+1}^{(1)}.$$

We can now renormalise the projection operators $C\sb{i+1}\sp{(j)}$
and  define
\begin{equation}
\label{equation:eij}
e_j^{(i)}\equiv [[i+1]]_{q} C_{i+1}^{(j)} \; ,
\end{equation}
where
$[[k]]_{q}\equiv \frac{q^{k/2}-q^{-k/2}}{q^{1/2}-q^{-1/2}}
=q^{-(k-1)/2}[k]_q$.
The elements $e_j^{(i)}$ of $H\sb n(q)$ have the following properties
\begin{equation}
\label{equation:esq}
\left(e_j^{(i)}\right)^2= [[i+1]]_{{q}} e_j^{(i)}
\end{equation}
and
\begin{equation}
\label{equation:ep}
e_j^{(i+1)}=\frac{1}{[[i]]_{{q}}[[i+1]]_{{q}}}
\bigg(e_j^{(i)}e_{j+1}^{(i)}e_j^{(i)}-e_j^{(i)}\bigg)\; .
\end{equation}

Finally, we can use the automorphism
$g_{j+k}\mapsto g_{j+i-k}\; ,\;\;\; k=0,\, 1,\, \cdots,\, i,$
in $H_{i+2}^{(j)}(q)$, under which $C_{i+2}^{(j)}$ is invariant,
 to show that
\begin{equation}
\label{equation:eau}
e_j^{(i)}e_{j+1}^{(i)}e_j^{(i)}-e_j^{(i)}=
e_{j+1}^{(i)}e_{j}^{(i)}e_{j+1}^{(i)}-e_{j+1}^{(i)}\; .
\end{equation}

Equations (\ref{equation:eau}) ensure that the $e\sb j\sp{(i+1)}$
can be equivalently defined by interchanging indices $j$ with $j+1$
on the right hand side of (\ref{equation:ep}).
We also notice that equation (\ref{equation:eau}) is of the same form
as the
second equation in (\ref{equation:Hecke}) written in terms of the
generators $e\sb i$ (\ref{equation:defe}). Indeed, for $i=1$,
$e\sb j\sp{(1)} = e\sb j$ and thus (\ref{equation:eau}) together
with (\ref{equation:esq}) and (4) are the defining
relations of the Hecke algebra $H\sb n(q)$. One  should notice,
however, that for general $i$, $e\sb j\sp{(i)} e\sb{j+k+1}\sp{(i)}
\neq e\sb{j+k+1}\sp{(i)}e\sb{j}\sp{(i)}$ if $k = 1, \ldots, i-1$.
Hence, the $e\sb{j}\sp{(i)}$ do not generate any Hecke algebra for
$i>1$. But for $i=n-2$ there are only two elements $e\sb 1\sp{(n-2)}$,
$e\sb 2\sp{(n-2)}$ and thus they generate the Hecke algebra
$H\sb 3(\tilde{q})$ where $\tilde{q}$ is determined by the equation
$[[2]]\sb{\tilde{q}} = [[n-1]]\sb q$.

\section{The Temperley-Lieb-Martin algebras}
In this section we derive the algebraic relations defining
the quotient of the Hecke algebra corresponding to
irreducible representations with at most $\ell$ columns. To do
so  we start from $H_{\ell+1}(q)$,
the lowest order Hecke algebra for which such restriction is
meaningful. Since the only irreducible representation of
$H_{\ell+1}(q)$ on which
$e_1^{(\ell)}$ does not vanish is $[\ell+1]$, the desired
quotient corresponds to $e_1^{(\ell)}=0$.

We can now state the main result of this article.
\begin{thm}
In the Hecke algebra $H_n(q)$, let $e_i^{(\ell)}$, where
$\ell=1,\, 2,\, \cdots,\, n-1$ and $i=1,\, 2,\, \cdots,\, n-\ell$,
be given by eq. (\ref{equation:eij}). Then the following are equivalent:
\begin{enumerate}
\item
$e_i^{(\ell)}=0\; ; \; i=1,\, 2,\, \cdots,\, n-\ell.$
\item
$H\sb n(q)$ is restricted to have  irreducible
representations labelled by Young diagrams with at most $\ell$ columns.
\end{enumerate}
\label{theorem.acropolis}
\end{thm}
\proof
The Hecke algebra $H_n(q)$, $n\geq \ell+1$, can be written as a
direct sum of the three subalgebras
$H_i^{(1)}(q)$, $H_{\ell+1}^{(i)}(q)$ and
$H_{n-i-\ell+1}^{(i+\ell)}(q)$, where the first and/or the last
could be the trivial algebra $H_1(q)$.
Therefore, the irreducible representations of $H_n(q)$ are direct
products of the irreducible representations of
the three Hecke subalgebras specified.
To show that 1 follows from 2 we note that
if only irreducible representations with at most $\ell$ columns are
allowed for $H_n(q)$,
then by Corrolary~\ref{cor.product}  only such irreducible
representations are allowed for each of the
subalgebras.
In particular, for the $H_{\ell+1}^{(i)}(q)$ algebra the
irreducible representation $[\ell+1]$ is excluded.
Consequently, $e_i^{(\ell)}$
inevitably vanishes ({\it cf.} eq. (\ref{equation:eij})).

To show that 2 follows from 1 we note that from 1
$\tr(e_1^{(\ell)})=0$.
Recall that in $H_{\ell+1}(q)$ $\tr(e_1^{(\ell)})$ vanishes on
all irreducible representations with not more than $\ell$ columns and
is positive on
the irreducible representation $[\ell+1]$.
Given any irreducible representation $\Gamma_n$ of $H_n(q)$, $n>\ell+1$,
the trace of $e_1^{(\ell)}$
can be evaluated recursively via \cite{Katriel2}
$$\tr(e_1^{(\ell)})_{\Gamma_n}^{\phantom{G_n}}=
\sum_{\Gamma_{n-1}\subset\Gamma_n}
\tr(e_1^{(\ell)})_{\Gamma_{n-1}}^{\phantom{G_n}}\; ,$$
where $\Gamma_{n-1}\subset\Gamma_n$ means that $\Gamma_{n-1}$
is one of the Young diagrams
obtained by eliminating a box in $\Gamma_n$. Now,
if $\Gamma_n$ consists of more than $\ell$ columns it means that
the iterative process carries the positive contribution initially
due to $\tr(e_1^{(\ell)})_{[\ell+1]}^{\phantom{G_n}}$ and that
cannot be annuled since
there are no negative contributions. Hence, for all but
irreducible representations with at most $\ell$ columns
$\tr(e_1^{(\ell)})>0$.
\endproof

Noting that for $\ell =2$ the first statement of
Theorem~\ref{theorem.acropolis}
is simply the relation (\ref{equation:bra}) that defines the \tl
algebra we have the following well known
\begin{cor} {\mbox{(\cite{Jones87,Good})}}
The \tl algebra $TL\sb n(q)$ is a quotient of the Hecke algebra
$H\sb n(q)$
admitting only irreducible representations corresponding to Young
diagrams with not more than two columns.
\end{cor}
Using eq. (3) it is a simple matter to reduce the
condition obtained for  $\ell =3$, {\it i.e.,} $e_i^{(3)}=0$,
to the form derived by Sochen in \cite{Sochen}.
\begin{rk}
{\rm From the proof of Theorem~\ref{theorem.acropolis} it follows that
the requirement $e\sb{1}\sp{(\ell)}=0$ is sufficient to eliminate all
Young diagrams with more than $\ell$ columns.
Once this is established it is a simple matter to show that
$e_i^{(\ell)}=0$ for all $i\leq n-\ell$. This can be done either
by invoking the other half of  Theorem~\ref{theorem.acropolis}
(the fact that $e_i^{(\ell)}=0$ is necessary for the irreducible
representations to contain at most
$\ell$ columns), or by noting that $\tr(e_i^{(\ell)})=\tr(e_1^{(\ell)})$
\cite{Katriel2}, and that $e_i^{(\ell)}$ is central in
$H_{\ell}^{(i)}(q)$,  i.e. within any irreducible representation
it is a multiple of the unit matrix.}
\label{remark}
\end{rk}

Trivial modifications yield the multi-row
Temperley-Lieb algebra. Precisely, we introduce the projection
operator
$$R_i=\prod_{j=2}^i \frac{L_j-q}{(-q)[j]_{q^{-1}}} \; ,$$
that annihilates all Young diagrams with $i+1$ boxes in which
the box $(2,1)$, whose $q$-content is $q$, is present. Thus,
the sole surviving Young diagram is the
single-column (i.e. $(i+1)$-row) diagram $[1^{i+1}]$.
Instead of eq. (\ref{equation:eij}) we obtain
$$f_j^{(i)}=[[i+1]]_{\sqrt{q}} R_{i+1}^{(j)}\; .$$
In particular, for $i=1$ we obtain
$$f_i=q^{\frac{1}{2}}(q-g_i) \, .$$
Eqs. (\ref{equation:esq}) and (\ref{equation:ep}) remain
unchanged in form, except that $f_j^{(i)}$ replaces $e_j^{(i)}$.
In Theorem~\ref{theorem.acropolis} ``columns'' should be replaced
by ``rows''.

\section{Elimination of rectangular subdiagrams}

In this section we define the quotient of the Hecke algebra that
corresponds to
Young diagrams which do not contain a
rectangular subdiagram  consisting of $\ell_v$ rows
each of length $\ell_h$. We start by considering the lowest
order Hecke algebra for which such a restriction is meaningful,
$H_{\ell_h\ell_v}(q)$.
Let
$$C_{\ell_h}=\frac{\prod_{i=2}^{\ell_h\ell_v}
\Big(L_i-q[\ell_h]_q\Big)}
{\prod_{i=1}^{\ell_h}\prod_{j=1}^{\ell_v}{\phantom{,}}^{\prime}
(q[j-i]_q-q[\ell_h]_q)}$$
be the projection operator that eliminates diagrams in which
the top box in the column $\ell_h+1$ is occupied,
and
$$R_{\ell_v}=\frac{\prod_{i=2}^{\ell_h\ell_v}
\Big(L_i - q[-\ell_v]_q\Big)}
{\prod_{i=1}^{\ell_h}\prod_{j=1}^{\ell_v}{\phantom{,}}^{\prime}
(q[j-i]_q-q[-\ell_v]_q)}$$
be the projection operator that eliminates diagrams in which the
leftmost box in the row $\ell_v+1$ is occupied. The prime
in both expressions means that the product in the denominator
excludes the factor $i=j=1$.
$C_{\ell_h}$ and $R_{\ell_v}$ are normalised to unity on the
rectangular diagram $[\ell_h^{\ell_v}]$.

Since $R_{\ell_v}C_{\ell_h}$ vanishes on all irreducible
representations except
$[\ell_h^{\ell_v}]$, the quotient of interest is specified by
setting $Q_{\ell_h,\ell_v}\equiv R_{\ell_v}C_{\ell_h}=0$.

If an  irreducible representation of $H_n(q)$,
$n>\ell_h\ell_v$, does not contain the
rectangular subdiagram $[\ell_h^{\ell_v}]$ we find that
$$Q_{\ell_h,\ell_v}^{(i)}\equiv R_{\ell_v}^{(i)}C_{\ell_h}^{(i)}=0
\; ,$$
in all subalgebras $H_{\ell_h\ell_v}^{(i)}$,
$i=1,\, 2,\, \cdots,\, n+1-\ell_h\ell_v$. Here
$R_{\ell_v}^{(i)}$ and $C_{\ell_h}^{(i)}$ are obtained from
$R_{\ell_v}$ and $C_{\ell_h}$, respectively, by replacing $L_j$
by $L_j^{(i)}$.

On the other hand, if $Q\sb{\ell_h,\ell_v}^{(1)}=0$ then in
all irreducible representations of $H_n(q)$, $n>\ell_h\ell_v$,
$\tr ( Q_{\ell_h,\ell_v}^{(1)}) =0$.
In an irreducible representation $\Gamma_n$ that contains
the rectangle $[\ell_h^{\ell_v}]$
$\tr(Q_{\ell_h,\ell_v}^{(1)})>0$, since the recursive
evaluation of this trace carries the positive contribution of
$\tr(Q_{\ell_h,\ell_v}^{(1)})_{[\ell_h^{\ell_v}]}$.
Recall that $[\ell_h^{\ell_v}]$ is
the only irreducible representation of $H_{\ell_h\ell_v}(q)$
for which $\tr(Q_{\ell_h,\ell_v}^{(1)})\neq 0$. Therefore
if we require that $Q\sb{\ell_h,\ell_v}^{(1)} =0$ no
representation containing $[\ell_h^{\ell_v}]$ is allowed.

Thus, the vanishing of all $Q_{\ell_h,\ell_v}^{(i)}$,
$i=1,\, 2,\, \cdots,\, n+1-\ell_h\ell_v$, is a necessary and sufficient
condition for the exclusion of irreducible representations that contain
the rectangular subdiagram $[\ell_h^{\ell_v}]$.

As an example we consider the exclusion of $[2^2]$,  i.e.
$\ell_h=\ell_v=2$, in $H\sb 4(q)$.
Here,
$$Q_{2,2}^{(1)}=\frac{(L_2-(q+q^2))(L_3-(q+q^2))(L_4-(q+q^2))
(L_2+1+\frac{1}{q})(L_3+1+\frac{1}{q})(L_4+1+\frac{1}{q})}
{(-q^2)(-1-q-q^2)(-q-q^2)(q+1+\frac{1}{q})
(\frac{1}{q})(1+\frac{1}{q})}\; .$$
In fact, $L_2$ can only assume the eigenvalues $q$ and $-1$,
corresponding to the boxes $(1,2)$ and $(2,1)$, respectively.
Therefore, the factors containing $L\sb 2$ in the numerator do not
annihilate any Young diagram and appear to be superfluous.
To write a normalised projector in a simpler form
we define the projectors $Q_s$ and $Q_a$, onto the two
vectors spanning the $[2,2]$ irreducible representation,  i.e.
$[2][2,1][2,2]$ and $[1,1][2,1][2,2]$, respectively.
Thus,
$$Q_s=\frac{(L_2+1)(L_3-q-q^2)(L_4-q-q^2)(L_4+1+\frac{1}{q})}
{(1+q)(-1-q-q^2)(-q-q^2)(1+\frac{1}{q})}$$
and
$$Q_a=\frac{(L_2-q)(L_3+1+\frac{1}{q})(L_4-q-q^2)(L_4+1+\frac{1}{q})}
{(-1-q)(q+1+\frac{1}{q})(-q-q^2)(1+\frac{1}{q})} \; .$$
Clearly, $Q_s+Q_a$ is equal to unity within the irreducible
representation $[2,2]$,
and vanishes otherwise.

In \cite{Martin} a different condition was proposed
to exclude the $[2,2]$ irreducible representation of $H\sb 4(q)$.
Namely, it was required that the operator
$$Q_{mr}= e\sb 1 e\sb{3}e\sb{2}([[2]]\sb q - e\sb 1)
([[2]]\sb q - e\sb{3})\; ,$$
should vanish. It can easily be checked, however, that  $Q\sb{mr}$
is not a projection operator and furthermore it
is nilpotent, i.e., $(Q\sb{mr})^2 = 0$.

\section{Conclusions}
In this paper we have presented a simple method for constructing
quotients of the Hecke algebra $H\sb n(q)$, the irreducible
representations of which are labelled by Young diagrams with
suitably restricted shapes. Our construction is based on the use of
the Murphy operators in $H\sb n(q)$. In particular, we
have shown that the $\ell$-column
\tlm algebra is defined by the relations
\begin{equation}
e_j^{(\ell)}e_{j\pm 1}^{(\ell)}e_j^{(\ell)}=e_j^{(\ell)},
\quad j = 1, \ldots, n-\ell
\label{equation:tl.generalised}
\end{equation}
which have a form identical with the \tl relations (\ref{equation:bra}).

The restriction of Hecke algebras to representations corresponding
to Young diagrams of a given shape have been
investigated in \cite{Martin} in order to analyse the spectra of
hamiltonians
of integrable models with $U\sb q(su(N,M))$ symmetry such as
the Perk-Schultz  quantum chains, and also to analyse a
certain class of diffusion-reaction processes \cite{Alcaraz}. Other
restrictions were used in classifaction of conformal field
theories \cite{Sochen}.
We believe that our construction and, in particular,
the simple form (\ref{equation:tl.generalised}) of the relations
defining the appropriate
quotients can be used in further analysis of such models.

\vspace{1cm}

\noindent
{\bf Acknowledgements}

We would like to thank Professor C. Quesne for helpful discussions.
TB is grateful to the European
Union Commission for the HCM fellowship. His work was also
partially supported by the grant KBN 2 P 302 21706 p 01.
JK's reaserch was carried out in the framework of a project
supported by the Fonds National de la Recherche Scientifique.


\end{document}